\documentclass[runningheads]{llncs}
\usepackage{graphicx}

\usepackage[cmex10, fleqn]{amsmath}
\usepackage{amssymb}
\usepackage{cite}
\usepackage{subfig}
\usepackage{graphicx}
\usepackage{booktabs}
\usepackage{listings}
\usepackage[dvipsnames]{xcolor}
\usepackage{array}
\usepackage{mdwtab}         
\usepackage{eqparbox}
\usepackage{algorithm}
\usepackage{algorithmic}
\usepackage{xspace}
\usepackage{url}
\usepackage{tikz}
\usepackage{wrapfig, blindtext}

\pdfmapfile{+zi4.map}
\usepackage{zi4}
\spnewtheorem{thm}{Example}[section]{\bfseries}{\itshape}

 \definecolor{colKeys}{rgb}{0,0,1}
 \definecolor{colIdentifier}{rgb}{0,0,0}
 \definecolor{colComments}{rgb}{1,0,0}
 \definecolor{colString}{rgb}{0,0.5,0}
 
\newcommand\Tstrut{\rule{0pt}{2.6ex}}         %
\newcommand\Bstrut{\rule[-0.9ex]{0pt}{0pt}}   %

\newcommand{\nli}[1]{\\\hspace{#1mm}}

\newcommand{\bindent}[1]{%
  \begingroup %
  \setlength{\itemindent}{#1} %
  \addtolength{\algorithmicindent}{#1} %
}
\newcommand\eindent{\endgroup} %

\newcommand*\circled[1]{\tikz[baseline=(char.base)]{
            \node[shape=circle,draw,inner sep=0.5pt] (char) {#1};}}
\begin{document}
\title{Static Detection of Uninitialized Stack Variables in Binary Code}
\titlerunning{Static Detection of Uninitialized Stack Variables in Binary Code}
\author{Behrad Garmany \and
Martin Stoffel \and
Robert Gawlik \and
Thorsten Holz
}
\authorrunning{Garmany et al.}
\institute{Horst Görtz Institute for IT-Security (HGI)\\ 
Ruhr-Universität Bochum, Germany\\
\email{\{firstname.lastname\}@rub.de} }
\maketitle              %
\begin{abstract}
More than two decades after the first stack smashing attacks, memory
corruption vulnerabilities utilizing stack anomalies are still prevalent and
play an important role in practice. Among such vulnerabilities, uninitialized
variables play an exceptional role due to their unpleasant property of
unpredictability: as compilers are tailored to operate fast, costly
interprocedural analysis procedures are not used in practice to detect such
vulnerabilities. As a result, complex relationships that expose uninitialized
memory reads remain undiscovered in binary code. Recent vulnerability reports
show the versatility on how uninitialized memory reads are utilized in
practice, especially for memory disclosure and code execution. Research in
recent years proposed detection and prevention techniques tailored to source
code. To date, however, there has not been much attention for these types of
software bugs within binary executables.

In this paper, we present a static analysis framework to find uninitialized
variables in binary executables. We developed methods to lift the binaries
into a knowledge representation which builds the base for specifically
crafted algorithms to detect uninitialized reads. Our prototype
implementation is capable of detecting uninitialized memory errors in complex
binaries such as web browsers and OS kernels, and we detected 7 novel bugs.
\end{abstract}
\section{Introduction}

Memory corruption vulnerabilities are prevalent in programs developed in
type-unsafe languages such as C and C++. These types of software faults are
known since many years and discovering memory corruption bugs in binary
executables has received a lot of attention for decades. Nevertheless, it is
still an open research problem to efficiently detect vulnerabilities in binary
code in an automated and scalable fashion. Especially \emph{temporal} bugs seem to be a
common problem in complex programs, an observation that the steady stream of
reported vulnerabilities confirms~\cite{cve_code_ex_ie, cve_code_ex_fx,
cve_code_ex_chrome, Szekeres:2013:EWM}.  In practice, especially web browsers
are often affected by temporal bugs and these programs suffer from
use-after-free vulnerabilities, race conditions,
uninitialized memory corruptions, and similar kinds of software vulnerabilities. 

One specific challenge is the efficient detection of uninitialized memory errors, such as uninitialized
stack variables. While such vulnerabilities got into the focus of several
real-world attacks~\cite{CVE_2014_6355, CVE_2015_0061, cve_code_ex_ie, CVE_2012_1889, pwn2own16, pwn2own17}, they
still represent an attack vector that is not studied well and often
overlooked in practice. The basic principle of such vulnerabilities is
straightforward: if a variable is declared but not defined (i.e., not initialized properly) and used later on in
a given program, then an attacker may abuse such a software fault as an attack
primitive. The uninitialized variable may for example contain left-over
information from prior variables in stale stack frames used during prior function calls. 
This information can be used to disclose memory and
leak sensitive information, which can then be used by an attacker 
to bypass \emph{Address Space Layout Randomization} (ASLR) or other defenses. In the worst case, an attacker
can control the content of an uninitialized variable and use it to execute
arbitrary code of her choice, hence fully compromising the program.
Uninitialized memory errors represent a vulnerability class that often affects complex, real-world programs: for
example, at the 2016 edition of the annual \emph{pwn2own} contest, Microsoft's
\emph{Edge} web browser fell victim to an uninitialized stack variable~\cite{pwn2own16}. As a
result, this vulnerability was enough to exploit the memory corruption vulnerability and gain full control
over the whole program. Similarly, an uninitialized structure on the stack was used in the \emph{pwn2own} 
contest 2017 to perform a guest-to-host privilege escalation in VMware~\cite{pwn2own17}.

The detection of uninitialized variables in an automated way has been studied
for software whose source code is
available\cite{khurshid2003generalized,goanna,horwitz1995demand}. 
The urge for such systems, especially targeting the stack, is also addressed by recent research through tools like \textsc{SafeInit}~\cite{bos:safeinit} or \textsc{UniSan}\cite{UniSan}. These systems set their main focus on prevention and also rely on source code.

In practice, however, a lot of popular software is unfortunately proprietary
and only available in binary format. Hence, if source code is unavailable, we
need to resort to binary analysis. The analysis of binary code, on the  other hand,
is much more challenging since some of the context
information gets lost during the compilation phase. 
The loss of data and context information (e.g, names, types, and structures of data are no longer available) hampers analysis and their reconstruction 
is difficult~\cite{dde:ndss11, jin2014recovering, mempick:wcre13}.
Thus, the development of precise analysis methods is more complicated
without this information. Addressing this issue, we are compelled to consider every statement in the assembly code as it
might relate to uninitialized memory of stack variables.

In this paper, we address this challenge and propose an automated analysis
system to statically detect uninitialized stack variables in binary code.
Since dynamic analysis methods typically lack comprehensive coverage of all
possible paths, we introduce a novel static analysis approach which
provides full coverage, at the cost of potentially unsound results (i.e.,
potential false positive and false negatives). However, unveiling potential
spots of uninitialized reads and covering the whole binary poses a more
attractive trade-off given the high value of detecting novel vulnerabilities.
Note that the information obtained by our approach can further serve in a
dynamic approach, e.g., to automatically verify each warning generated by our
method.

Our analysis is performed in two phases: First, we designed a framework to
lift binary software into an intermediate representation, which is further
transformed into a knowledge base that serves our \emph{Datalog} programs. We opted
for Datalog given that this declarative logic programming language enables us
to efficiently query our deductive database that contains facts about the
binary code. Based on Datalog, we then devised an accurate points-to analysis
which is both flow- and field-sensitive. More specifically, with points-to
information we have explicit information on indirect writes and reads that are
connected to passed pointers. This allows us to track the indirect read or
write back to the specific calling context where the points-to information is
further propagated and incorporated in our knowledge base. This analysis step
builds up the conceptual structure on which our dataflow algorithms to detect 
uninitialized variables operate.

To demonstrate the practical feasibility of the proposed method, we
implemented a prototype which is tailored to detect 
uninitialized stack variables. Our results show that we can successfully 
find all uninitialized stack vulnerabilities in the Cyber Grand Challenge (CGC) binaries.
In addition, we detected several real-world vulnerabilities in complex
software such as web browsers and OS kernel binaries. Finally, our prototype
is able to detect and pinpoint new and previously unknown bugs in programs
such as \emph{objdump}, and \emph{gprof}.

\smallskip \noindent
In summary, our main contributions in this paper are:

\begin{itemize}

\item We design and implement an automated static analysis approach and introduce several processing steps which enable us to encode the complex data flow within a given binary executable to unveil unsafe zones in the control flow graph (i.e., basic blocks in which potential uninitialized reads might occur).
 
\item We present a flow-, context- and field-sensitive analysis approach built on top of these processing steps, suitable for large-scale analysis of binary executables to detect uninitialized reads in a given binary executable.

\item We evaluate and  demonstrate that our analysis framework can detect both
vulnerabilities in synthetic binary executables and complex, real-world
programs. Our results show that the framework is capable of finding new bugs.

\end{itemize}

\section{Uninitialized Stack Variables}
Stack variables are local variables stored in the stack frame of a given
function. A function usually allocates a new stack frame during its prologue
by decreasing the stack pointer and setting up a new frame pointer that
points to the beginning of the frame. Depending on the calling convention,
either the caller or callee take care of freeing the stack frame by
increasing the stack pointer and restoring the old frame pointer. For
example, in the \texttt{stdcall} calling convention, the callee is
responsible for cleaning up the stack during the epilogue. It is important to
note that data from deallocated stack frames are \emph{not} automatically
overwritten during a function's prologue or epilogue. This, in particular, means
that old (and thus stale) data can still be present in a newly allocated
stack frame. A stack variable that is not initialized properly hence contains
old data from earlier, deallocated stack frames. Such a variable is also
called \emph{uninitialized}. An uninitialized stack variable can lead to
undefined behavior, not at least due to its unpleasant property that the
program does not necessarily crash upon such inputs. In practice,
uninitialized variables can be exploited in various ways and pose a serious
problem~\cite{CVE_2014_6355, CVE_2015_0061, cve_code_ex_ie, CVE_2012_1889,
pwn2own16,nuern:uinit}. They usually contain junk data, but if an attacker
can control these memory cells with data of her choice, the software
vulnerability might enable arbitrary code execution.

To tackle this problem, the compiler can report uninitialized stack variables
at compile time for intraprocedural cases. Unfortunately, interprocedural
cases are usually not taken into account by compilers. This lies in the
nature of compilers which need to be fast and cannot afford costly analysis
procedures. Even for optimization purposes, past research reveals that the
benefits of extensive interprocedural analyses are not large enough to be
taken account of in compilers~\cite{Richardson:1987}.
\section{Design}

In the following, we provide a comprehensive overview of our static analysis
framework to detect uninitialized stack variables in binary executables. Our
analysis is divided into two processing stages. In a pre-processing step, we
lift the binary into an IL and transform each function into SSA with respect
to registers. The transformed functions are translated into Datalog facts
which serve as our extensional database or \emph{knowledge base}.

Based on this database, we then perform an interprocedural
points-to analysis with respect to stack pointers. The analysis also results
in information about pointers that are passed as arguments to functions and hence we can
determine those pointers that enter a new function context.
The reconstructed information about indirect definitions and uses of stack
locations is used in a post-processing state in which we determine \emph{safe
zones} for each stack access. Safe zones consist of \emph{safe basic blocks},
i.e., a use in these blocks with respect to the specific variable is covered by
a definition on all paths. Stack accesses outside their corresponding safe
zone produce warnings. For each variable, we determine a safe zone in its
corresponding function context.

\begin{figure*}[tb]
	\centering
    \includegraphics[width=0.8\textwidth]{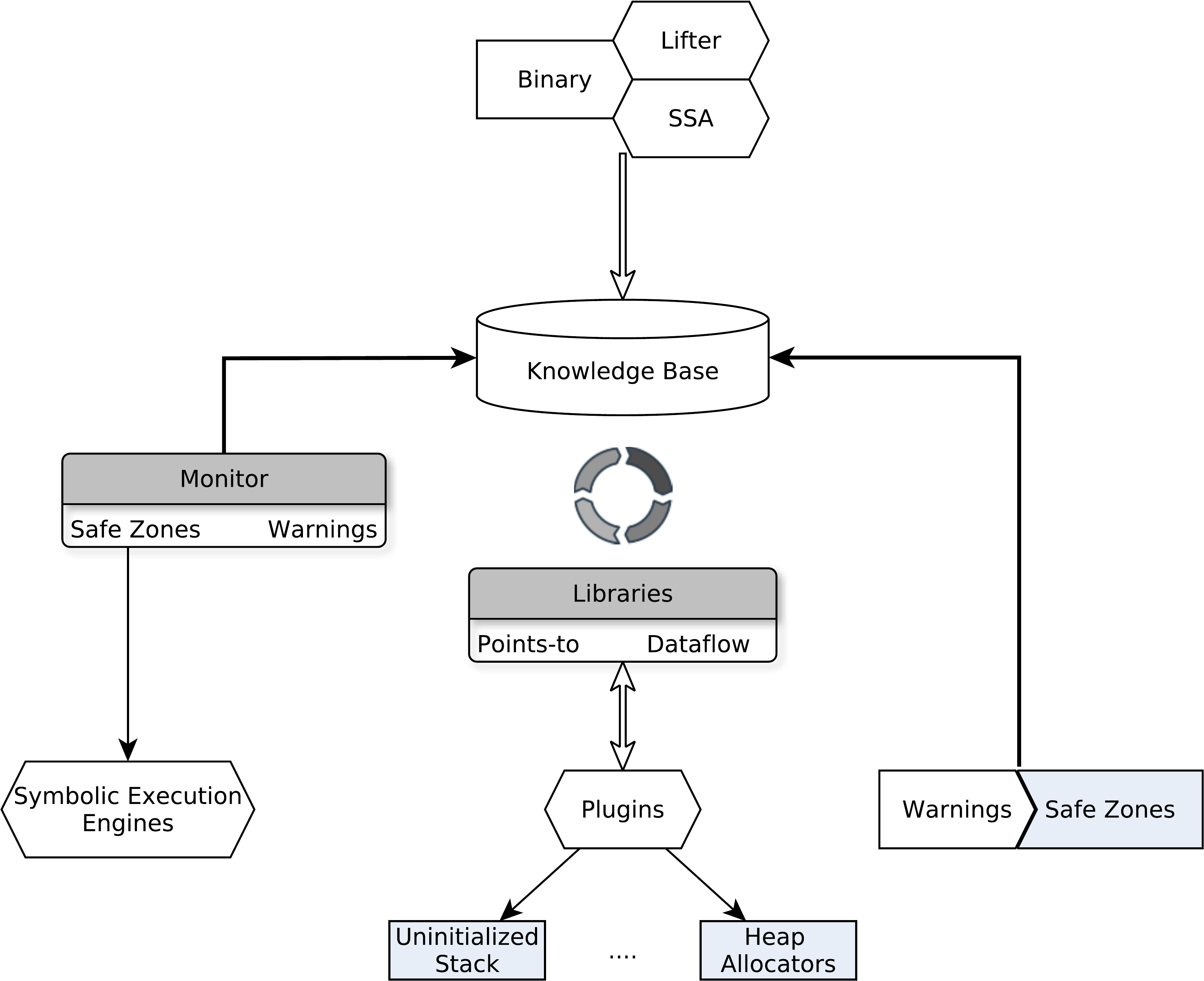}
    \caption{Architecture implemented by our prototype.}
    \label{fig:arch}
\end{figure*}

Our dataflow algorithms propagate information about safe zones from callers
to callees and vice versa. If a path exists from the entry of a function to
the use of a variable, which avoids the basic blocks of its safe zone, then
the stack access is flagged \emph{unsafe}. If a potentially uninitialized parameter is passed to a function, we check if
the exit node of the function belongs to its safe zone. This in particular
means that each path from the entry point of the function to the leaf node is
covered by a definition (i.e., initialization) of the variable. We propagate this information back
to the call sites, i.e., the fallthrough basic block at the call site is
added to the safe zone of the variable in the context of the caller. This
information in turn is further propagated and used at other call sites.
Figure~\ref{fig:arch} shows the architecture implemented by our prototype.
Our design allows to attach different checker plugins into the system that
work and enrich the same knowledge base with valuable information. Each
plugin can be run in parallel. Whenever a new information enters the
knowledge base, the plugins adapt to it. 

Warnings and safe zones are outputs of the analysis phase and put into the knowledge base.
A monitor observes changes made to safe zones and warnings to either spawn the 
Datalog algorithms or a symbolic execution engine. The symbolic execution engine
tackles the path sensitivity and is fed with safe zones of each stack variable. The
aim of the symbolic execution engine is to reach the warning, i.e., a potential
uninitialized read, by avoiding the safe zone of that variable. 
The whole procedures cycle, i.e., each component contributes to the knowledge base
which in turn is consumed by other components to adapt.

Our current prototype is tailored towards uninitialized stack variables.
However, as the plugin system suggests, we are able to enrich the analyses
with heap information (see \S~\ref{sec:heapallocs}).
In the next sections, we explain the individual analysis steps in detail and
present examples to illustrate each step.
\subsection{Stack Pointer Delta}
\label{sec:spd}

\begin{figure}[tb]
    \centering
    \includegraphics[scale=1.0]{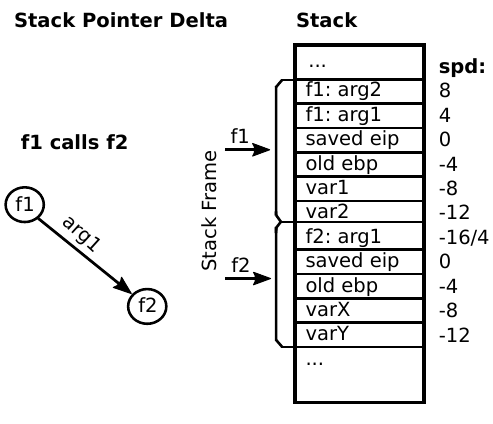}
    \caption{Stack Pointer Delta (spd) addressing.}
    \label{fig:spd_example}
\end{figure}

On Intel x86 and many other instruction set architectures, the stack pointer is used to keep
a reference to the top of the stack. The stack is usually accessed in
relation to the current stack pointer. On Intel x86, the frame pointer is
additionally linked to the stack pointer and keeps a reference to the
beginning of the current stack frame. A function stack frame can, therefore,
be located on different stack addresses depending on its stack pointer value.
Because of this, we can not refer to a stack address directly. Instead, we
use a delta value depending on the stack pointer. We refer to this addressing
type as \emph{stack pointer delta addressing}. A stack frame always starts at
delta zero and grows towards negative values. Therefore, local variables are located at 
negative offsets, while passed arguments through the stack reside at positive offsets. We handle
the \emph{fastcall} calling convention for both x64 and x86 architectures in
a generic way by means of Datalog rules . 
To simplify the analysis, we rebase all memory accesses on the stack to be
relative to the stack pointer instead of using the frame pointer.
\begin{definition}
	\label{def:stack_variable}
	Let $S$ be the set of all stack variables. Each $s \in S$ is a tuple of the form $(spd_s, fld_s)$ where $spd_s$ is the 
	stack pointer delta of $s$ and $fld$ an optional field value added to the base address of the underlying memory object.
\end{definition}
Figure~\ref{fig:spd_example} illustrates this concept. Function \texttt{f1}
calls \texttt{f2}. The resulting stack is shown on the right-hand side. In
each stack frame the saved EIP value is mapped to the delta value zero.  
A direct access to var1 inside function \texttt{f1} is associated with $(-8, 0)$, since it is 
at delta $-8$. Variable \emph{varX} is also at delta value $-8$ but inside function
\texttt{f2}. Each delta value is associated with the corresponding function in which the access
occurs. 

\subsection{Points-To Analysis for Binaries}
\begin{figure}[thb!]
\noindent\begin{minipage}{.50\textwidth}
	\scriptsize
	~\\[-5.3em]
\begin{algorithmic}
	\STATE Input: $EDB$ 
	\STATE Output: Points-To Facts
	\STATE
	\STATE\circled{1} VPtsTo(V1, SPD, Addr, Ctx) $\leftarrow$ 
	   \bindent{2.5em}
		   	\STATE StackPointer(V1, Addr, SPD).
	   	\eindent
	\STATE
	\STATE\circled{2} VPtsTo(V1, SPD, Addr, Ctx) $\leftarrow$
       \bindent{2.5em}
       		\STATE Assign(V1,V2, Addr),
       		\STATE VptsTo(V2, SPD, \_, Ctx).
		\eindent
	\STATE
	\STATE\circled{3} VPtsTo(V1, SPD2, Addr, Ctx) $\leftarrow$
	  \bindent{2.5em}
	   		\STATE Load(V1, V2, Disp, Addr, Ctx),
			\STATE VPtsTo(V2, SPD, Addr2, Ctx),
			\STATE CanReach(Addr, Addr2, Ctx),
			\STATE PointerPtsTo(SPD, Disp, SPD2).
	  \eindent
	\STATE
	\STATE\circled{4} PointerPtsTo(SPD,Disp,SPD, Ctx) $\leftarrow$
		  \bindent{2.5em}
		   		\STATE Store(V1, Disp, V2, Addr, Ctx),
				\STATE VPtsTo(V1, SPD, \_, Ctx),
				\STATE VptsTo(V2, SPD, \_, Ctx).
		  \eindent
		
	\STATE
	\STATE\circled{5} VPtsTo(Res, SPD+Value, Addr, Ctx) $\leftarrow$ 
		  \bindent{2.5em}
				\STATE BinOp(Op, Res, V1, V2, Addr, Ctx),
				\STATE VPtsTo(V1, SPD,  \_, Ctx),
				\STATE Constant(V2, Value, Addr).
		  \eindent
	\STATE
	\STATE\circled{6}  VPtsTo(V2, SPD2, CalleeAddr, Callee) $\leftarrow$
				  \bindent{2.5em}
					\STATE Param(V1, Arg, Addr, Caller, 
					  \bindent{5.8em}
						\STATE V2, CalleeAddr, Callee),
					   \eindent
				  	\STATE TranslateSPD(Arg, Callee, SPD2),
				  	\STATE VptsTo(V1, \_, \_, Caller).
				  \eindent
	\STATE
	\STATE\circled{7} VPtsTo(V1, SPD, Addr, Ctx) $\leftarrow$
			 \bindent{2.5em}
			 	\STATE Phi(V1, PhiReg, Addr, Ctx),
			 	\STATE VptsTo(V1, SPD, \_, Ctx).
			 \eindent
	\STATE
	\STATE\circled{8} IndirectDef(V1, SPD, Addr) $\leftarrow$
						  \bindent{2.5em}
						    \STATE Store(V1, Disp, V2, Addr, Ctx),
						  	\STATE VptsTo(V1, SPD, \_, Ctx).
						  \eindent
				
	\STATE
		\STATE\circled{9} IndirectUse(V1, SPD, Addr) $\leftarrow$
				  			\bindent{2.5em}
							    \STATE Load(V1, V2, Disp, Addr, Ctx),
							  	\STATE VptsTo(V1, SPD, \_, Ctx).
                               \eindent
\end{algorithmic}
\end{minipage}
\hfill%
\begin{minipage}[t]{.49\textwidth}
	\scriptsize
	\noindent\begin{tabular}{l p{3.7cm}}
	\textbf{Variables:}\Tstrut\Bstrut & \nli{2}
		$Addr$\Tstrut  					  & Instruction address\nli{2}
		
		$V_{i}$\Tstrut 					  & Register or memory expression\nli{2}
		
		$SPD$\Tstrut				      & Stack pointer delta value (\emph{spd}) of the corresponding stack location\nli{2}
		
		$Ctx$\Tstrut\Bstrut   		      & Function name/context\nli{2}
		
		$Arg$\Tstrut 					  & Part of Param facts; describes which parameter we are dealing with $(1st, 2nd, \ldots)$ \\
	\hline
	\textbf{EDB Facts:}\Tstrut\Bstrut & \nli{2}
	
	StackPointer 		         & A fact that unifies V1 with the register that holds a pointer. The Addr of the instruction and the \emph{spd} value of the stack location are stored in the corresponding variables, respectively.\nli{2}
	
	Assign\Tstrut				 & Corresponds to a \texttt{mov} instruction in assembly\nli{2}
	
	Load\Tstrut				 	 & Dereference from $V_2 + Disp$ and store it into $V_1$\nli{2}
	
	Store\Tstrut                 & Store content of $V_2$ into memory at $V_1+Disp$\nli{2}
	
	Param\Tstrut				 & Describes a parameter pass at the call site $(Addr)$ from actual $V_1$ in the caller context to a formal $V_2$ in the callee context\nli{2}
	TranslateSPD\Tstrut 		 & Translates the \emph{spd} value of the parameters in the context of the callee and vice versa. In x86 first parameter has \emph{spd} value 4, second has 8 etc.\nli{2}
	
	BinOp\Tstrut 				& Describes a binary operation where Op is applied on $V_1$ and $V_2$ and the result is stored in $Res$.\nli{2}
	
	Constant\Tstrut 		    & Describes a constant value used in a binary operation\nli{2}
	
	Phi\Tstrut 					& Corresponds to an SSA \emph{phi} assignment. $PhiReg$ is bound to registers in the \emph{phi} expression
	\end{tabular}
\end{minipage}\\[0.5em]
\hfill%
\begin{minipage}[t]{.45\linewidth}
    \captionof{algorithm}{Points-To Analysis.}\label{alg:points-to}
\end{minipage}%
\hfill%
\begin{minipage}[t]{.45\linewidth}
	\captionof{figure}{Variables and EDB facts.}\label{fig:tableEDB}
\end{minipage}%
\end{figure}
Analyzing binary machine code poses many challenges to overcome. The lack of
type information forces us to analyze each IL statement. To overcome the
complicated arrangement of work list algorithms, we decided to opt for a
declarative approach. Rather than solving the problem by an imperative
algorithm, we describe the problem and let the solver perform the evaluation
strategy. Therefore, we utilize an Andersen-style algorithm~\cite{Andersen94}
in Datalog which is flow- and field-sensitive. Our algorithm is inspired by
recent research done by Smaragdakis et
al.~\cite{Smaragdakis:2010,Smaragdakis:2015}. They show how context, flow,
and field sensitivity can be achieved in large-scale through a Datalog-based
approach. We adapted their approach and tailored our algorithms for binary
analysis.

Each information about memory loads, stores, assignments, arithmetic
operations, control flow, and parameter passes which is expressed in terms of
the IL, is extracted into an extensional database (\emph{EDB}).
For each binary, an EDB is produced which represents a knowledge base of
a priori facts.

A simplified version of our approach delivers the idea and is presented by
Algorithm~\ref{alg:points-to}. The Datalog algorithm is fed with the EDB which
builds the base for Datalog rules to derive new facts. These new facts 
build the intensional database (\emph{IDB}). The \emph{IDB} and \emph{EBD}
form our \emph{knowledge base}.
In Figure~\ref{fig:tableEDB}, we summarize the facts and variables used in
Algorithm~\ref{alg:points-to}. Some rules are left-out for the sake of
brevity.

\paragraph{\textbf{Datalog.}} To better understand what Algorithm~\ref{alg:points-to}
does, we refer the reader to common literature on logic programming. 
Datalog is a declarative logic programming language that
has its roots in the database field~\cite{ceri} and its general purpose is to
serve as a query language for large, complex databases. Datalog programs run
in polynomial time and are guaranteed to terminate. Conventional Datalog uses
a Prolog-like notation, but with simpler semantics and without the data
structures that Prolog provides. Its approach to resolve new facts is close
to what dataflow algorithms do with arrangements of worklist algorithms. It
strives a set-oriented approach which we require, rather than a goal-oriented
approach as done in Prolog.
A Datalog program consists of facts and rules which are represented as
\emph{Horn clauses} of the form: $P_0 :- \quad P_1, \ldots ,P_n$, where $P_i$
is a literal of the form $p(x_1,\ldots,x_k)$ such that $p$ is a predicate and
the $x_j$ are terms. The left-hand side of the clause is called \emph{head};
the right-hand side is called \emph{body}. A clause is true when each of its
literals in the body are true.

\subsubsection{Rules.}
Refer to Algorithm~\ref{alg:points-to}. The predicate \texttt{VPtsTo} stands for
the points-to set of a variable.
Rule~\circled{2} specifies the following: Given an assignment
from \texttt{V2} to \texttt{V1} at a specific address, include the points-to
set of \texttt{V2} into that of \texttt{V1}.
\noindent Rule~\circled{5} specifies a case of derived pointers: Given a binary
operation such that \texttt{Res = V1 + V2}, where \texttt{V1} is a pointer,
check if \texttt{V2} is a constant. If the conditions hold, then \texttt{Res}
points to a stack location with a stack pointer delta of \texttt{SPD+Value}.
Variable \texttt{Value} is grounded by the third fact in the body of this
rule.
With rule~\circled{5} we deduce a new points-to set that
corresponds to a new stack location, which in turn is again used to derive new
points-to set information by the recursive chain of rules. This procedure
is performed on all rules until the process saturates, i.e., a fixpoint is reached where we do not
gain additional facts.
Another rule that deserves attention is rule~\circled{6}. Here points-to
information is tracked into the context of the callee. The rule specifies
that if a stack pointer is passed as a parameter, then a new points-to set is
created for that parameter in the context of the callee. If, for example, a
stack pointer is passed as an argument, then \texttt{TranslateSPD} in the
body of the rule gives us its stack pointer delta value in the context of the
callee. A new points-to set is created for this argument which is basically
a new derived fact. This fact, in turn, falls into the recursive chain to
serve for deducing new facts. The procedure provides an on-demand approach to
track arguments of a function, only when they are passed by reference. We
achieve context-sensitivity by introducing tags at each call sites. These
tags are linked with the parameters.

To illustrate the approach, let's assume that our analyzer runs over the following piece of code:
\begin{lstlisting}[language={[x86masm]Assembler}, 
				   keywordstyle=\bfseries, basicstyle=\ttfamily\scriptsize]
	foo:                                     main:
	  0x8049000 mov eax, [esp+4]              ...				  											
	  0x8049004 mov dword [eax], 0xff         0x80490f0: lea ebx, [esp-0x30] 
	  0x804900a mov [eax+4], eax              0x80490f4: push ebx 
	                                          0x80490f5: call foo
\end{lstlisting}
At \texttt{0x80490f0-0x80490f4} a stack pointer is pushed onto the stack with a delta of \texttt{-0x30} which
resides in \texttt{[esp+4]} in the context of \texttt{foo}. As the result of the preprocessing step, we have
\texttt{Param} and \texttt{TranslateSPD} facts extracted into our EDB (see
Figure~\ref{fig:tableEDB}). For this example, we have the facts
\texttt{Param([esp\_17], 1, 0x80490f5, "main", [esp+4], "foo")}, and \texttt{TranslateSPD(1,
"foo", 4)}, where \texttt{[esp\_17]} corresponds to the location the stack pointer is pushed to at \texttt{0x80490f4}.
Since we are dealing with a passed stack pointer, our analysis
derives \texttt{VptsTo([esp\_17], -0x30, 0x80490f4. "main")}. 
By using rule \circled{6}, the points-to analysis can now deduce
the fact \texttt{VPtsTo([esp+4], 4, 0x8049000, "foo")}. 
Note that the pointer is now considered to have a delta of $4$ in the \emph{new context}. We do this
to keep track of pointers that are parameters, otherwise we lose focus on where the pointer might 
\emph{originate} from. 

With rule \circled{2}, we get the connection to \texttt{eax}, and with rule
\circled{3}, \circled{4} the connection of \texttt{[eax+4]} to the
underlying memory object, i.e., a pointer to itself. By Definition~\ref{def:stack_variable}, 
we refer to \texttt{[eax+4]} as $(4,4)$ since the base points to a location with delta 4
and we are accessing the location that is 4 bytes apart from the base.
\subsection{Safe Zones}
\label{sec:sz}
\begin{figure}[thb!]
\noindent\begin{minipage}{.60\textwidth}
	\scriptsize
\begin{algorithmic}[1]
	\STATE Input: $\textsc{CFG}=(V,E),\,\textsc{Defs},\,\textsc{Uses}$
	\STATE $\quad\quad\;\, \textsc{ DOM}\,(dominator \, sets)$
	\STATE Outputs: \textsc{SafeZone}, $E$ (\textsc{SafeEdges})\\[1ex]
	\STATE Let $E'=\textsc{SafeEdges} = \{\}$
	\STATE Let $\textsc{SafeZone} = \textsc{Defs}$\\[1ex]
	\STATE Let \textsc{Vars} = $\bigcup\limits_{(spd,fld,bb)\, \in \,\textsc{Defs}\, \cup \,\textsc{Uses}} (spd,fld)$\\[1ex]
		\FORALL{$(spd, fld, bb) \in \textsc{SafeZone}$}
			\STATE $E' = E' \cup \{(spd, fld, bb, bb_x) \,|\, (bb,bb_x)\in E\}$
			\Tstrut
			\FORALL{$bb_d \in \textsc{DOM}(bb)$}
				\STATE $E' = E' \cup \{(spd, fld, bb_d, bb_x) \,|\, (bb_d,bb_x)\in E\}$
				\STATE $\textsc{SafeZone} = \textsc{SafeZone} \cup \{(spd,fld,bb_d)\}$
				\Tstrut
			\ENDFOR	
		\ENDFOR

	\STATE Let \textsc{Unsafe} = $\{\}$
		\FORALL{$(spd,fld,bb) \in \textsc{Vars}$} 
		\STATE \textbf{if} $\exists p=<bb_{start},\ldots,bb_i, bb_j,\ldots,bb>$ with 
		\STATE $\;\;\;(spd,fld,bb_i, bb_j) \notin E', \forall bb_i, bb_j \in p, i\neq j$
		\STATE \textbf{then} $\textsc{Unsafe} = \textsc{Unsafe} \cup \{(spd,fld,bb)\}$
		\STATE \textbf{else} $\textsc{SafeZone} = \textsc{SafeZone} \cup \{(spd,fld,bb)\}$
		\ENDFOR
	\captionof{algorithm}{Sketch: Computation of Safe Zones}\label{alg:safezone}
\end{algorithmic}
\end{minipage}
\begin{minipage}{.5\textwidth}
    \includegraphics[scale=0.20]{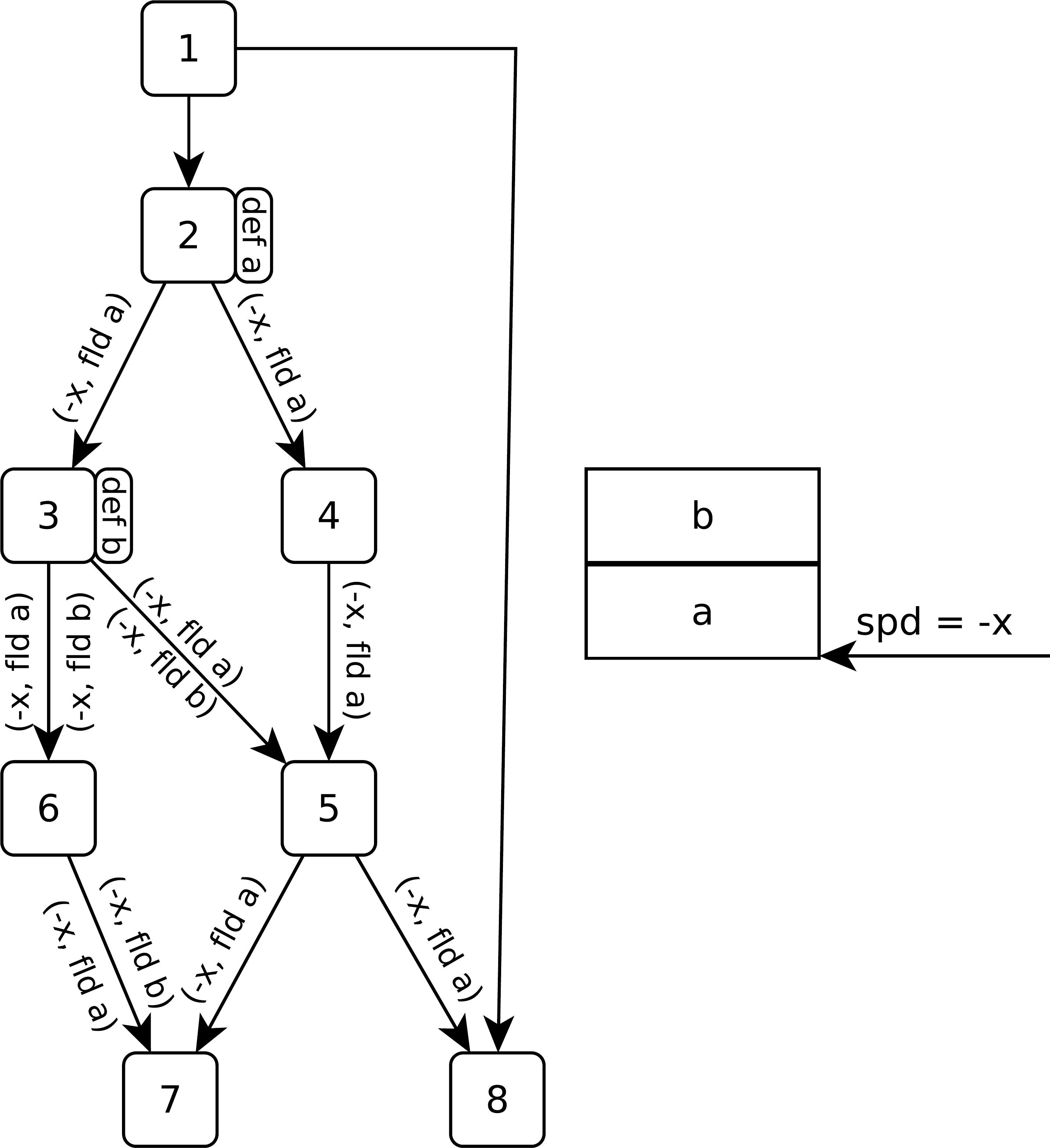}
    \captionof{figure}{Graphical representation  \\of labeling safe edges.}\label{fig:safeunsafe}
\end{minipage}
\end{figure}

In this section we describe our approach to determine if a given stack read
is safe. With \emph{safe} we refer to the property that a read is covered by
its definitions on all paths. Each basic block where a \emph{safe} read
occurs is considered a \emph{safe basic block}. Since we are dealing with
different memory objects, the set of safe basic blocks is different for each
object/variable. More formally, we define it as follows.
\begin{definition}
	\label{def:safe_bb}
	Let $CFG = (V,E)$ be the control flow graph, S the set of all stack variables, and let $Defs=\{(spd,fld,bb_s) \,|\, bb_s \in V, (spd,fld) \in S\}$ be the set
	of all stack accesses that define the stack location $(spd, fld)$ at $bb_s$. $bb_s$ is called a safe basic block.
	Each edge that originates from $bb_s$ is called a safe edge with respect to $(spd, fld)$. Each safe edge is a tuple of the form
	$(spd, fld, bb_s, bb_t)$ with $(bb_s,bb_t) \in E$. The set of all safe basic blocks with respect to $(spd, fld)$ is called the safe zone of $(spd, fld)$.
\end{definition}
Apparently, if all incoming edges to a basic block are \emph{safe edges} with respect to some variable \texttt{(spd,fld)}
then that basic block is a \emph{safe basic blocks} for that variable. 
To determine \emph{safe zones} of each variable we proceed as
sketched in Algorithm~\ref{alg:safezone}. An unsafe read occurs if a path
exists that avoids the safe zone, i.e., a path from the entry of the function
to the \emph{use} location which does not go through safe edges. Lines
$17-21$ generalize this procedure for all stack accesses. 
If such a path does not exist, we flag the basic block as safe. 

In its essence the algorithm does a reaching definition analysis for each
stack variable and labels the basic blocks and edges accordingly. The initial
process for building safe zones is achieved through lines $7-13$. From each
definition node with respect to the specific stack variable, the information
about its definition is propagated further along its path in the control flow
graph.

Each stack variable is associated with its own safe zone.
Note that we do not use memory SSA as it introduces conflicts and complicates 
the points-to analysis. Hence, benefits of SSA in this manner are marginal.
\begin{thm}
Figure~\ref{fig:safeunsafe} illustrates the labeling of safe edges. Basic
blocks \texttt{2} and \texttt{3} define variables $a$ and $b$, respectively.
Each use of the variable in these basic blocks is considered \emph{safe}.
Accordingly, each use of variable $b$ in $\{3,6\}$ is considered safe. At a
stack pointer delta of $-x$ an access to variable $a$ is possible. Its
field/offset ($fld\;a$) is zero. For an access to variable $b$, $fld\;b$ is
added. Safe Zone with respect to $(-x,fld\;a)$ consists of basic basic blocks
$\{2,3,4,6,5,7\}$. For $(-x, fld\;b)$ we have $\{3,6\}$. Each use of the
variable in these basic blocks is considered \emph{safe}.
\end{thm}
\subsection{Interprocedural Flow Analysis}
\label{sec:interproc}
During the data flow analysis, we propagate state information that concerns
the initialization of passed arguments between caller and callee. Information
about pointers is passed back and forth by the points-to analysis. We use
this information to determine indirect accesses (see rule
\circled{8},\circled{9} in Algorithm~\ref{alg:points-to}). If a leaf node in
the callee context is flagged \emph{safe} with respect to a stack access, we
flag the corresponding call sites as safe. This procedure propagates
information back to the caller, extending the safe zone in the caller
context. In turn, Algorithm~\ref{alg:safezone} needs another run by using the
new information and distinguish between unsafe and safe accesses to the
stack. Previously unsafe accesses might turn to safe accesses through this
process. This procedure is repeated until it saturates, i.e., no changes to
the set of safe basic blocks.

\noindent\textbf{Summaries:}
A common technique used in interprocedural static analysis is the use of summaries~\cite{Sharir}. 
These summaries can be block summaries which gather the effects of a basic block, 
or function summaries that gather the effects of the whole function with respect to
the variables of interest. Whenever a function call is encountered, these summaries  
are utilized and applied. The facts in Datalogs EDB and its deduced facts through rules
in the IDB can be seen as such summaries. Whenever a function call is encountered, the
analysis uses facts about the function that concern the variables of interest.

\noindent\textbf{Multiple analyses plugins:}
As shown in Figure~\ref{fig:arch}, our design has a plugin mechanism. All plugins
operate on the same knowledge base. 
Plugins deduce and incorporate knowledge into the knowledge base which
can transparently be tapped by other plugins and library routines.
This, for instance, allows the \emph{Uninitialized Stack} plugin to operate on
information deduced by the \emph{Heap Allocators} plugin.
Each plugin can be run in parallel and whenever new 
information enter the knowledge base, the plugins adapt to it. Each change to the 
knowledge base with respect to warnings and safe zones is monitored. 

\noindent\textbf{Detecting Uninitialized Accesses:}
When the analysis reaches its fixpoint, all information about safe and unsafe
zones with respect to all stack accesses is present. A stack access outside
its safe zone causes a warning. Additionally, we track each use to its origin,
i.e., in the case of a stack pointer, we track it to the call site where the
pointer originates from.

\subsection{Symbolic Execution}
For each warning, we need to check if a satisfiable path exists to the use of the variable by
avoiding its safe zone. We therefore need a mechanism for path-sensitivity in our analysis process.
To tackle path sensitivity, we utilize under-constrained symbolic execution~\cite{Ramos:2015}.
Under-constrained symbolic execution immensely improves the scalability by checking
each function directly, rather than the whole program. Due to the lack of context before the 
call of the function under analysis some register/memory values are unconstrained, hence the
under-constrained term.

For each variable that caused a warning, we feed the symbolic execution
engine with information about its safe zones. Satisfiability is checked from
the function entry to the flagged variable by avoiding the basic blocks in
its safe zone. 
We start at each origin, i.e., the function where the stack variable
originates from. To improve the scalability of the symbolic execution, we
initially skip each function call. If a path is satisfiable then we might
have an overapproximation, since some skipped function might have made a
constraint become unsatisfiable. 

For unsatisfiable paths, we look at the \emph{unsat-core}, i.e. those
constraints which have caused the path become unsatisfiable. A function that
alters one of these variables in those constraints is then set free to be
processed by the engine; again in a similar fashion by first skipping calls
in the new function context until we eventually reach a satisfiable state.
The only difference is that we now force the engine to run into basic blocks
that modify the variables that made our constraints become unsatisfiable.
As a result, we basically overapproximate the set of satisfiable paths.
Filtered warnings are removed as such in the knowledge base.

\section{Implementation}
\label{sec:impl}

Our prototype is implemented on top of the Amoco framework~\cite{amoco}.  The
decision for Amoco is favored due to its flexible IL. It allows us to extend its
set of expressions. Each new  expression transparently integrates and
interplays with the standard set of expressions.

We retrieve the control-flow graph from the disassembler IDA Pro which is
shown to be the most accurate\cite{andriesse2016}. Each basic block is
transformed into an Amoco basic block. We extended Amoco to support SSA and
implemented the algorithm proposed by Cytron et al.~\cite{cytronSSA}. In
particular, we adapted the concept of collectors that are described in Van
Emmerik's work on decompilers~\cite{emmerik}. Collectors can be seen as an
instrumentation of the SSA algorithm. The algorithm proposed by Cytron et al.
uses a stack of live definitions whenever a basic block is processed. This
information is valuable to put into a knowledge base. For instance, we can
instrument the algorithm to write a set of live definitions at call sites
into our knowledge base which we use to translate SSA subscripted expressions
back and forth between caller and callees. Due to SSA with respect to
registers, we obtain partial flow sensitivity. 

We built the symbolic execution on top of angr~\cite{YAN:2016}, a platform agnostic binary
analysis framework. As Figure~\ref{fig:arch} indicates, we plan to attach more engines to 
our framework. This is motivated by the fact that each engine comes with advantages and 
its shortcomings, which we hope to compensate by combining different engines. 
The points-to analysis results are saved into a separate database. If
extensions are needed, we can reuse this database and let Datalog evaluate new
facts based on the new extensions.

\section{Evaluation}
\label{sec:eval}
\begin{table*}[tb]
\scriptsize
\caption{Analysis results for the relevant CGC binaries that contain an
uninitialized memory vulnerability.}
\centering
\label{tab:eval1}
\renewcommand\tabcolsep{5pt}
	\begin{tabular}{c  c  c  c  c  c }
		\toprule
		\textbf{Binary} & \textbf{Functions} & \textbf{Facts} & \textbf{Pointer Facts} &\textbf{Stack\;Accesses} & \textbf{Unique\;Warnings}\\
		\midrule
		Hackman	   & 70	  & 46k   &   545  &  943   & 9   \\
		Accel 	   & 185  & 109k  &   2179 &  2057  & 33  \\
		TFTTP      & 58	  & 30k   &	  175  &  600   & 3   \\
		MCS        & 122  &	156k  &   860  &  2498  & 11  \\
		NOPE       & 105  & 57k	  &   568  &  1378  & 8   \\
		Textsearch & 90	  &	50k	  &   290  &   597   & 2  \\
		SSO        & 64	  & 26k	  &   204  &   650   & 4  \\
		BitBlaster & 10	  & 4k	  &   42   &   95    & 1  \\
		\bottomrule
	\end{tabular}
\end{table*}

In this section, we evaluate the prototype implementation of our analysis framework 
and discuss empircal results. Note that the analysis framework is by design {OS independent.
Our analyses were performed on a machine running with Intel~Xeon~CPUs
\textsc{E5-2667@2.90GHZ} and \textsc{128GB RAM}. The programs presented in 
Table~\ref{tab:eval1} and Table~\ref{tab:eval2} are  compiled for \emph{x86-64}. 
Our prototype is not limited to 64~bit, but also supports 32~bit binaries.

\subsection{CGC Test Corpus}

As a first step to obtain a measurement on how our approach copes with
realistic, real-world scenarios, we evaluated our prototype over a set of
Cyber Grand Challenge (CGC) binaries which, in particular, contain a known
uninitialized stack vulnerability. These CGC binaries are built to imitate
real-world exploit scenarios and deliver enough complexity to stress out
automated analysis frameworks. Patches of the vulnerabilities ease the effort
to find the states of true positives and hence these binaries can serve as a
ground truth for our evaluation. We picked those binaries from the whole CGC
corpus that are documented to contain an uninitialized use of a stack variable
as an exploit primitive and we evaluate our prototype with these eight
binaries.

Table~\ref{tab:eval1} shows our results for this CGC test setup. 
The third column of the table depicts the number of facts extracted from the binary
building up the EDB. The fourth column shows the number of deduced pointer
facts. The fifth column depicts the total number of stack accesses. The sixth
column denotes the number of potential uninitialized stack variables grouped
by their stack pointer delta value and their origin. This approach is similar
to \emph{fuzzy stack hashing} as proposed by Molnar et al.~\cite{Molnar:2009}
to group together instances of the same bug.

For each of the eight binaries, we successfully detected the vulnerability.
Each detected use of an uninitialized stack variable is registered, among
which some might stem from the same origin. Therefore, we group those
warnings by the stack pointer delta values of those stack variables from
which they originate. The individual columns of Table~\ref{tab:eval1} depict
this process in numbers. We double-checked our results with the patched
binaries to validate that our analysis process does not produce erroneous
warnings for patched cases. For each patched binary, our analysis does not
generate a warning for the vulnerabilities anymore.

\subsection{Real-World Binaries}

\begin{table*}[tb]
\scriptsize
\caption{Analysis results for binutils-2.30, ImageMagick-6.0.8, gnuplot 5.2
patchlevel 4. Number in parentheses denotes the number of verified bugs.}
\label{tab:eval2}
\centering
\renewcommand\tabcolsep{5pt}
	\begin{tabular}{c  c  c  c  c  c}
		\toprule
		\textbf{Binary} & \textbf{Functions} & \textbf{Facts} & \textbf{Pointer Facts} &\textbf{Stack\;Accesses} & \textbf{Unique\;Warnings}\\
		\midrule
		objdump	   & 2.5k   & $>$4M    &   $>$19k  	&  23k  & 42\,(2)  \\
		ar 	   	   & 2.4k   & $>$3.2M  &   $>$16k  	&  19k  & 24       \\
		as-new     & 2.2k   & $>$2.9M  &   $>$9k   	&  15k  & 29       \\
		gprof      & 2.3k   & $>$3.8M  &   $>$16k  	&  20k  & 34\,(1)  \\
		cxxfilt    & 2.2k   & $>$3.1M  &   $>$15k   &  18k  & 22 	   \\
		ld-new     & 2.8k   & $>$3.8M  &   $>$16k  	&  22k  & 15       \\
		strings    & 2.2k   & $>$3.5M  &   $>$15k  	&  19k  & 11       \\
		size 	   & 1.9k   & $>$3.1M  &   $>$15k   &  19k  & 20       \\
		readelf    & 115    & $>$107k  &   $>$5k    &  543  & 4\,      \\
		gnuplot    & 3k     & $>$7.2M  &   $>$12    &  23k  & 54\,(2)   \\
		Image Magick & 6.5k & $>$24M   &   $>$31k   &  150k & 168\,(2) \\
		\bottomrule
	\end{tabular}
\end{table*}

Beyond the synthetic CGC test cases, we also applied our analysis framework
on real-world binaries. Table~\ref{tab:eval2} summarizes our results for
\emph{binutils} \emph{gnuplot}, and \emph{ImageMagick}. The values in
parentheses are manually verified bugs.
Note that the number of warnings pinpointing the root cause and the potential
flaw through an uninitialized variable is comparatively \emph{small} to the
number of all accesses. Additionally, our symbolic execution filter was able
to reduce the warning rate by a factor of eight in our experiments. Despite
the false positive rates which we discuss in the next sections, we strongly
believe that the output of our prototype is a valuable asset for an analyst
and the time spent to investigate is worth the effort.
The numbers are given in column~6 of Table~\ref{tab:eval2}. Overall, we found
two recently reported vulnerabilities in \emph{ImageMagick}, two previously
unknown bugs in \emph{objdump}, one unknown bug in \emph{gprof}, and two bugs
in \emph{gnuplot}. A manual analysis revealed that the bugs in objdump and
gprof are not security critical. The two bugs in gnuplot were fixed with the
latest patchlevel at the time of writing.

Our analysis can also cope with complex programs such as web browsers,
interpreters and even OS kernels in a platform-agnostic manner. We tested our
framework on \textit{MSHTML (Internet Explorer 8, CVE-2011-1346)},
\textit{Samba (CVE-2015-0240)}, \textit{ntoskrnl.exe (Windows kernel,
CVE-2016-0040)}, \textit{PHP (CVE-2016-7480)}, and \textit{Chakra (Microsoft
Edge, CVE-2016-0191)}. In each case, we successfully detected the vulnerability in
accordance to the CVE case.\\[2pt]

\noindent\textbf{Error Handling Code:}
Our study on hundreds of warnings shows that many warnings are internally
handled by the programs itself through error handling code. To address this
problem, we implemented a plugin that checks---starting from the corresponding
call site---if a return value might go into a sanitization process. We track
the dataflow into a jump condition and measure the distance to the leaf node.
If it is smaller than a threshold value of $3$, we assume that the return
value is sanitized and handled. This simple procedure works surprisingly well
in most cases to shift the focus away from paths that run into error handling
code.
\subsection{Heap Allocations} 
\label{sec:heapallocs}
User space programs use a variation of \texttt{malloc}  for  allocating memory
dynamically. Performance-critical applications like browsers even come with
their own custom memory allocators.  To enable tracking of dynamic memory
objects,  we use a list of known  memory allocators and enriched the knowledge
base with pointer information. The points-to analysis grabs this information
and deduces new facts. As a consequence we can observe a coherence between
stack locations and heap. While this is an experimental feature of our framework, it has proven itself
valuable by pinpointing three uninitialized bugs in \texttt{gprof} 
and \texttt{objdump} which originate from the heap.

\section{Discussion}
\label{sec:discussion}

The  discovery of vulnerabilities for both the CGC binaries and real-world
binaries demonstrates that our approach can successfully detect and pinpoint
various kinds of uninitialized memory vulnerabilities. 
Our analysis is tailored to stack variables, with a design that is well-aligned with
the intended purpose of a bug-detecting static analysis. 
However, it also comes with some limitations that are not currently tackled
by our prototype and we discuss potential drawbacks of our approach in the
following.\\[1pt]

\noindent\textbf{Heap:}
It is well known that analyzing the data flow on the heap is harder than data
flow on the stack and in registers. To address this problem, for example
Rinetzky and Sagiv proposed an approach to infer shape predicates for heap
objects~\cite{recursive_shape}. More recently, the topic of separation logic
has garnered more attention as a general purpose shape analysis
tool~\cite{separation_logic}. This is---among other reasons---du{}e to the
fact that aliasing analysis becomes much more difficult for real-world code
that makes use of the heap as compared to the data flow that arises from
stack usage. We account for all stack accesses under the reconstructed CFG.
Hence, a stack variable which is initialized through a heap variable is
supported by our approach. A points-to analysis needs to account for this
interplay. Therefore, we implemented a component that adapts to the given set
of points-to facts and tracks heap pointers originating from known heap
allocators~(see \S~\ref{sec:heapallocs}). This procedure is by design not
sound, however the discovered bugs which originated from the heap were found
by using this approach.

Many performance-critical applications like browsers have their own custom
memory allocators which poses a problem to address. However, there is work on
this field with promising results as shown in recent research done by Chen et
al.~\cite{Chen:2013, Chen:2016}.\\[1pt]

\noindent\textbf{False Positives/False Negatives:}
Many analyzers come with a set of strategies to deal with the number of
warnings by, for instance, checking the feasibility of paths. Each strategy
is usually tied to certain aspects of the problem, an approach which we
adapted and discussed in the last sections to tackle false positives.
However, we are dealing with many \emph{undecidable} problems here, i.e., the
perfect disassembly and the fact that detection of uninitialized variables
itself is \emph{undecidable} in general.

Aggressive compiler optimizations can also pose a burden on the
disassembly process as they can facilitate the problem of unpredictable control flow.
Even for a state-of-the-art tool like IDA Pro, control-flow analysis is
hampered by unpredictability and indirect jumps. Points-to information can
resolve some of these indirect jumps~\cite{Evans:2015}. However, its demand
for context sensitivity is expensive for large applications. Programs that
contain recursion further restrict the capabilities of static analysis, as
shown by Reps~\cite{Reps:2000}. A combination with dynamic approaches might
prove itself valuable and information derived by them can be incorporated
into our knowledge base. Recall, that each change in the knowledge base is
adapted transparently, i.e., facts are removed, added and deduced constantly
by adding or removing information from it.

A valuable feature for any analyzer is the question of \emph{soundness}. 
A fully sound analysis is hard to achieve in practice due to code that we
cannot analyze (e.g., libraries which are not modeled or dynamically
dispatched code where we might loose sight). A sound analysis needs a perfect
disassembly resulting in a perfect CFG, which is an \emph{undecidable} problem.
Therefore, false negatives are not avoidable. Similar source code based
systems use the term \emph{soundiness}~\cite{Livshits:2015:DSM} as they can
guarantee soundness for specific parts of the code under analysis
only~\cite{drchecker}. 

We strongly believe that if an analyzer finds enough errors to repay the cost
of studying its output, then the analyzer will be a valuable and
cost-effective instrument in practice.

\section{Related Work}
\label{sec:relwork}
The development of methods for static analysis spans a wide variety of
techniques. Modern compilers like the GNU-Compiler, MSVC, or Clang can report
uninitialized variables during compile time. They utilize the underlying
compiler framework to implement an intraprocedural analysis to detect
potential uninitialized variables. As discussed earlier, compilers are
trimmed to run fast, and the analysis time for costly interprocedural
algorithms is not desired. For optimization purposes, the benefits of
extensive interprocedural analyses might not be desirable to
apply~\cite{Richardson:1987}.

Flake was one of the first to discuss attacks against overlapping
data~\cite{delta_graphs}, an attack vector closely related to our work on
uninitialized memory reads. His presentation focuses on finding paths that
have overlapping stack frames with a target (uninitialized) variable.

Wang et al. present case studies about undefined behavior~\cite{Wang:2012}
among which are induced by the use of uninitialized variables. 
In a more recent work, Lee et al. introduce several methods to address
undefined behavior in the LLVM compiler with a small performance
overhead~\cite{Lee:2017:TUB}.

A popular attempt to tackle the problem of uninitialized memory reads is
binary hardening. \textsc{StackArmor}~\cite{bos:stackarmor} represents a hardening
technique that is tailored to protect against stack-based vulnerabilities. To
determine functions that might be prone to uninitialized reads, static
analysis is used to identify stack locations which cannot proven to be safe. 
The system can protect against uninitialized reads but cannot
detect them. \textsc{SafeInit}~\cite{bos:safeinit} extended this idea and represents a
hardening technique specifically designed to mitigate uninitialized read
vulnerabilities. The authors approach the problem from source code: based on
Clang and LLVM, the general idea is to initialize all values on the
allocations site of heaps and stacks. In order to keep the overhead low,
several strategies are applied to identify suitable spots. They modify the
compiler to insert their initialization procedures. By leveraging a
multi-variant execution approach, uninitialized reads can be detected. This,
however, needs a corpus of proper inputs that can trigger those spots.
\textsc{UniSan}~\cite{UniSan} represents a similar approach to protect operating
system kernels. Based on LLVM, the authors propose a compiler-based approach
to eliminate information leaks caused by uninitialized data that utilizes
dataflow analysis to trace execution paths that lead to possible leaking
spots. \textsc{UniSan} checks for allocations to be fully initialized when they leave
the kernel space, and instruments the kernel in order to initialize
allocations with zeros, if the check is violated.

Another recent approach by Lu et al.~\cite{nuern:uinit} targets uninitialized
reads in the Linux kernel. They propose
techniques for stack spraying to enforce an overlap of the sprayed data with
uninitialized memory. With a combination of symbolic execution and fuzzing,
they present a deterministic way to find execution paths which prepare data
that overlaps with data of a vulnerability.

Giuffrida et al.\cite{psi} present a monitoring infrastructure to detect
different kinds of vulnerabilities, among them uninitialized reads. They
perform static analysis at compile time to index program state invariants and
identify typed memory objects. The invariants represent safety constraints
which are instrumented as metadata into the final binary. Their approach also
allows to update and manage metadata dynamically. The proposed framework
monitors the application in realtime and checks for invariant violations. Ye
et al. propose a static value-flow analysis~\cite{Ye:vfg}. They analyze the
source and construct a value flow graph which serves to deduce a measure of
definedness for variables. The analysis results are used to optimize the
instrumentation process of binaries.

Other systems which instrument binaries either at compile or execution time
to detect uninitialized reads at runtime are proposed in the
literature~\cite{msan, drmemory}. These systems can be used in combination
with a fuzzer or a test suite to detect uninitialized variables. One
advantage of these dynamic systems is that for each detected uninitialized bug an
input vector can be derived. On the other hand, only executed paths will be
detected and hence the code coverage is typically low. In addition, an
appropriate corpus of input data is needed. In contrast, our static approach
is capable of analyzing binary executables in a scalable way that provides
high code coverage.

In summary, the wealth of work in recent research, most of
which rely on source code, is tailored to instrumentation purposes to aid
dynamic analysis in a monitoring environment. In contrast, our approach
follows a purely large-scale static analysis that addresses the proactive
detection of bugs in binary executables. 

\section{Conclusion}
\label{sec:conclusion}

Uninitialized memory reads in an application can be utilized by an attacker
for a variety of possibilities, the typical use case being an information
leak that allows an attacker to subsequently bypass information hiding
schemes. In this paper, we proposed a novel static analysis approach to
detect such vulnerabilities, with a focus on uninitialized stack variables.
The modularity of our framework enables flexibility. We have built a
prototype of the proposed approach that is capable of doing large-scale
analyses to detect uninitialized memory reads in both synthetic examples as
well as complex, real-world binaries. We believe that our system delivers new
impulses to other researchers.
\section*{Acknowledgements}

We thank the anonymous reviewers for their valuable feedback.
This work was supported by the German Research Foundation (DFG) within the
framework of the Excellence Strategy of the Federal Government and the States
-- EXC~2092 \textsc{CaSa} -- 39078197. In addition, this work was supported by
the European Research Council (ERC) under the European Union’s
Horizon 2020 research and innovation programme (ERC Starting
Grant No. 640110 (BASTION)).

\bibliographystyle{splncs04}
\bibliography{sigproc.bib}

\end{document}